\documentclass[a4paper]{spie}  

\usepackage{amsmath,amsfonts,amssymb}
\usepackage{graphicx}
\usepackage{framed}
\usepackage[colorlinks=true, allcolors=blue]{hyperref}
\usepackage{physics}		
\usepackage{nicefrac}
\DeclareMathOperator{\e}{e}	
\usepackage[euler]{textgreek}	
\usepackage{siunitx}
\usepackage[normalem]{ulem}

\newcommand{\mrm}{\mathrm}

\newcommand{\tre}{\tau_{\mrm{r}}}

\newcommand{\ts}{t_{\mrm{s}}}

\title{Trajectory control using an information engine}

\author{Tushar K. Saha}
\author{John Bechhoefer}
\affil{Simon Fraser University, Department of Physics, Burnaby, BC, Canada}

\authorinfo{Further author information: Send correspondence to TKS, tushars@sfu.ca, or JB, johnb@sfu.ca}

\begin{document} 
\maketitle

\begin{abstract}

We have built an information engine that can transport a bead in a desired direction by using favorable fluctuations from the thermal bath. However, in its original formulation, the information engine generates a fluctuating velocity and cannot control the position of the bead. Here, we introduce a feedback algorithm that can control the bead's position, to follow a desired trajectory. The bead can track the path if the maximum desired velocity is below the engine's maximum average velocity. Measuring the range of frequency that the feedback algorithm can track, we find a bandwidth that is slightly lower than the corner frequency of the bead in the trap.

\end{abstract}


\section{INTRODUCTION}
\label{sec:intro}  

Information engines are a new class of engines that can convert information about a system into useful energy. This class of engine was first imagined by Maxwell \cite{knott1911} more than 150 years ago. The considerable subsequent discussion in the literature to understand their operation has refined our understanding of the second law of thermodynamics.\cite{parrondo2015thermodynamics,leff2002maxwell} Recent technical advances have made it possible to realize these engines in the laboratory.\cite{toyabe2010,cottet2017,camati2016,koski2015chip,masuyama2018,chida2017power,kumar2018sorting, thorn2008, vidrighin2016} These technological developments include accurate and fast measurement of the state of microscopic systems and high computation speed, to apply feedback quickly. In the last decade, experiments have produced working Maxwell demons \cite{toyabe2010,koski2014} and tested the cost of operating the associated measuring device \cite{berut2012, jun2014, orlov2012, talukdar2017}. The focus of research has accordingly shifted from demonstrating the existence of the ``Maxwell demon" to understanding ways to maximize the performance~\cite{saha2021} and efficiency of these engines~\cite{Ritort2019, paneru2018, admon2018}. Experiments have also demonstrated related devices such as flashing ratchets~\cite{lopez2008,lau2020electron, skaug2018} and information ratchets\cite{serreli2007, ragazzon2018, alvarez2008}. In almost all of these studies, the focus was on the conversion of information into directed motion.  In Ref. \citeonline{saha2021}, we designed a ``useful'' information engine that can both store energy in a work reservoir and transport a bead at an average velocity in a desired direction.

Although our engine can control the average velocity, it cannot control the bead's position. As we will see below, random fluctuations lead to a diverging bead position with time. This observation is a consequence of the previously used ratcheting algorithms, which do not control the bead trajectory in space and time. In this article, we introduce an improved feedback algorithm to limit the position divergence and control the bead's trajectory. Trajectory control makes possible practical applications such as transporting a bead to a specific site at a particular time. Although other techniques such as photon nudging\cite{chicos2014}, fluid flow\cite{kumar2019fluid}, and magnetic steering \cite{carlsen2014} can also control the bead trajectory, such engines use conventional feedback algorithms that directly apply external forces to the bead. By contrast, information engines are powered by rectifying thermal fluctuations. No additional forces are applied to the bead, and no external work is done: The engine is powered purely by information. Here, we present a heuristic algorithm that can control the bead trajectory while respecting the no-work constraint, and we characterize the performance of the algorithm using techniques from control theory.\cite{Bechhoefer2021} We find that the performance of the engine is essentially limited by the engine's material properties and not by the algorithm. We also find that the force constraint in the information engine does not affect the tracking performance significantly when compared to driving the bead by directly applying large forces.

\section{INFORMATION ENGINE} 

\begin{figure} [ht]
   \begin{center}
   \includegraphics[width = 0.6\textwidth]{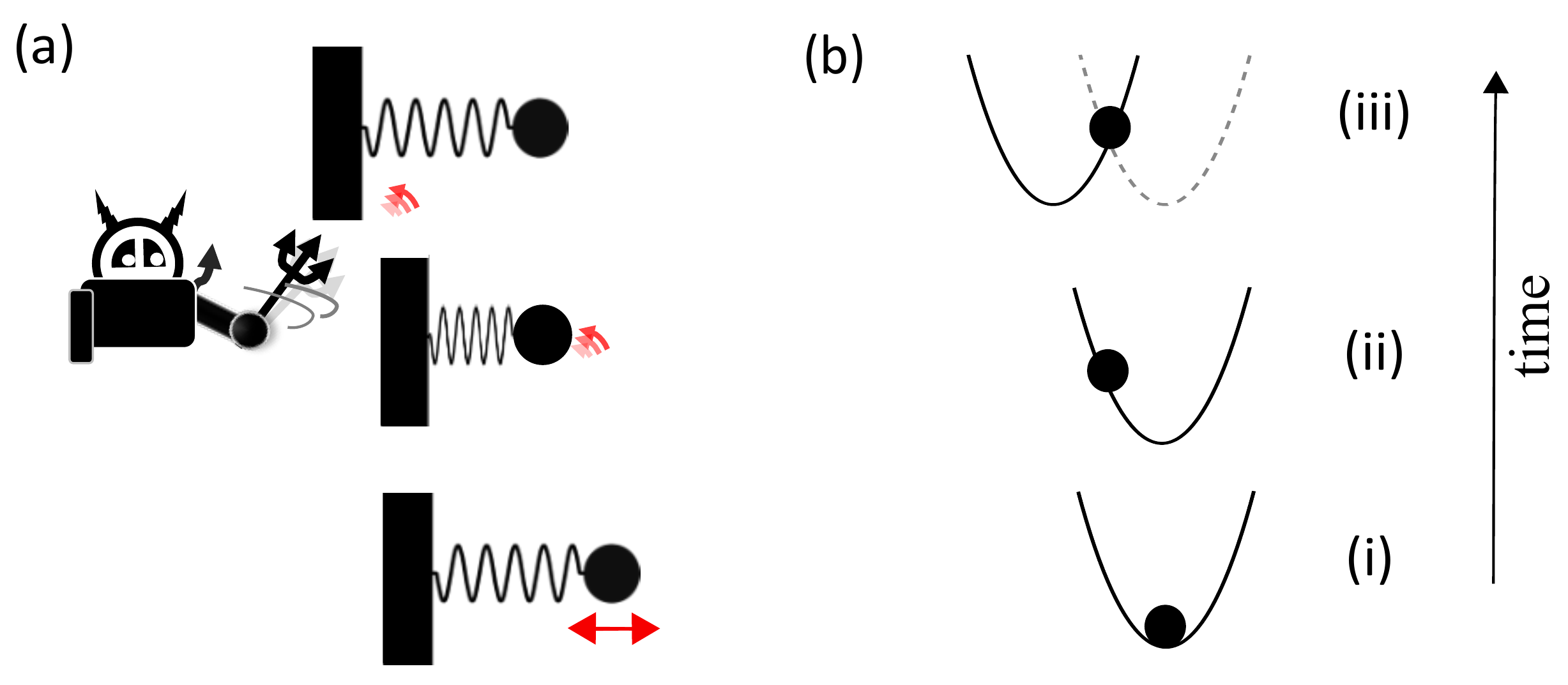}
   \end{center}
   \caption{ \label{fig:ratchet} 
Schematic diagram of the information-engine concept. (a) Ratcheted spring-mass system moves to the left; (b) equivalent dynamics in a harmonic well. }
\end{figure}

In previous work\cite{saha2021}, we developed an information engine that could lift an optically trapped bead at an average velocity against gravity. In this article, to simplify the discussion, we move the bead perpendicular to gravity to remove the effect of gravitational forces from the bead dynamics. Note that the same algorithm will also work for raising it against gravity. Conceptually, the engine can be viewed as a microscopic spring-mass system in contact with a thermal bath, as shown in Fig.~\ref{fig:ratchet}a. Because of the thermal noise, the mass fluctuates about an equilibrium position. Let the desired directed velocity be to the left.  Occasionally, the bead fluctuates left, compressing the spring. The favorable ``left" fluctuation is captured by immediately shifting the trap left, decompressing the spring. The corresponding energy of the spring-mass system is shown in Fig. \ref{fig:ratchet}b. The energy is visualized as a bead in a harmonic well. One important but subtle point is that the distance the spring (bottom of the harmonic well) is shifted should be chosen so that no external work is done by the spring on the bead. This is ensured by shifting the minimum of the well so that the bead is reset to the equi-potential state on the opposite side of the well, as shown in Fig.~\ref{fig:ratchet}b (iii). In effect, the information engine converts heat from the bath into directed motion.

\begin{figure} [ht]
   \begin{center}
   \includegraphics[width = 0.6\textwidth]{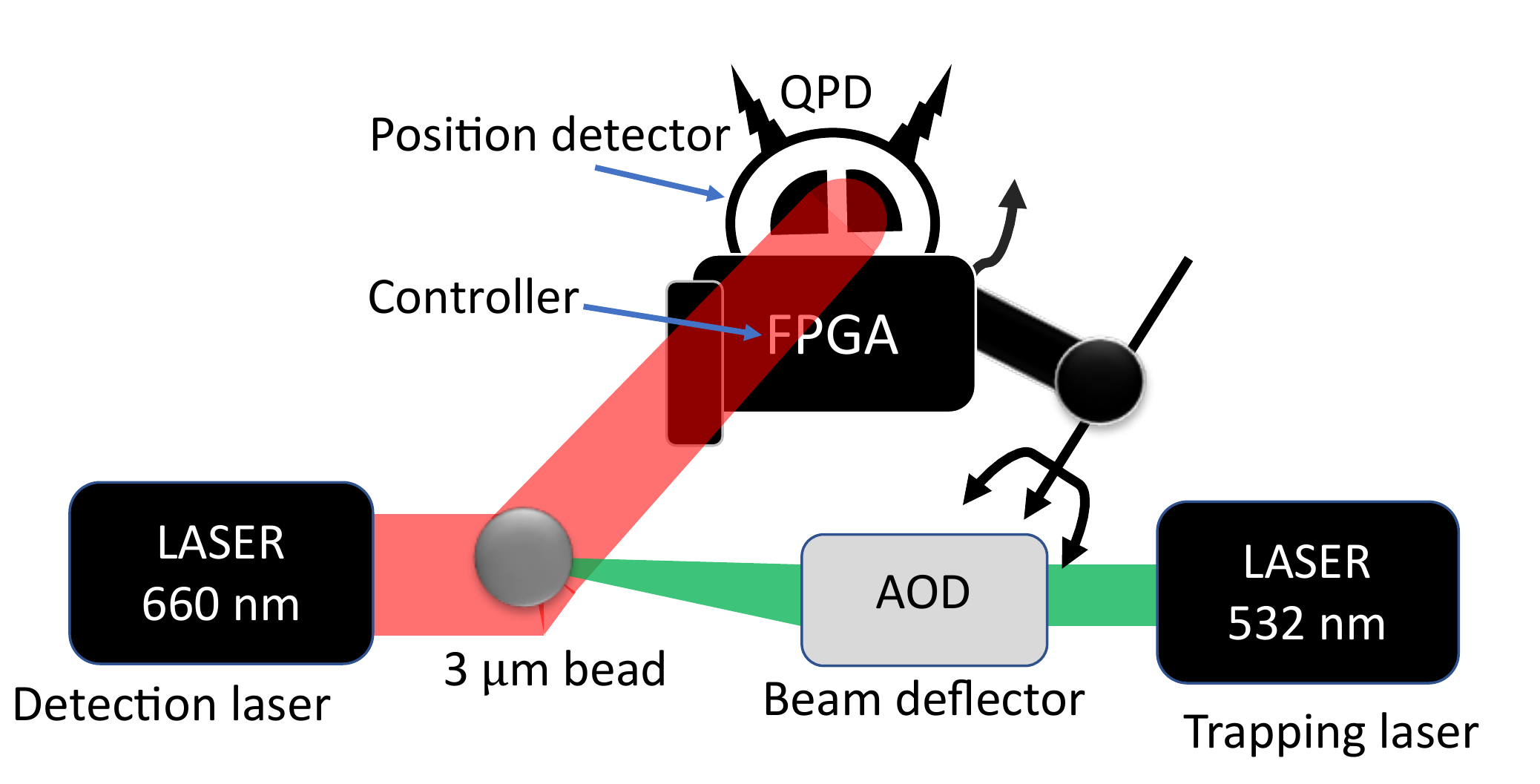}
   \end{center}
   \caption{ \label{fig:apparatus} Schematic diagram of the experimental set-up.
  AOD $=$ Acousto-Optic Defector, FPGA $=$ Field-Programmable Gate Array, and QPD $=$ Quadrant Photodiode. Representative components are not to scale. }
\end{figure}

Figure~\ref{fig:apparatus} shows the experimental set-up, which consists of a 3-\textmu m-diameter bead trapped by optical tweezers. Optical trapping is produced by a tightly focused 532-nm green laser (HÜBNER Photonics, Cobolt Samba, 1.5~W, 532 nm) using a microscope objective (60x Olympus water-immersion objective 1.2 NA). A 660-nm red laser (Thorlabs, LP660-SF20, 20 mW, 660 nm) is used as the detection laser by loosely focusing the laser beam on the bead using a second objective (40x Nikon, NA = 0.5). The two objectives are aligned to match their focal planes inside the sample chamber. Two acousto-optic deflectors (AODs, DTSXY-250-532, AA Opto Electronic) are used to steer the position of the trapping beam parallel and perpendicular to gravity. The red (660 nm) light collected from the trapping plane is focused on a quadrant photodiode (QPD) for position detection. The position of the bead is measured every 20 \textmu s (50 kHz) by the A/D converter of a Field Programmable Gate Array (FPGA, National Instruments, NI PCIe-7857), which calculates the feedback control signal response for the AOD. Further details on the experimental apparatus can be found in Ref.~\citeonline{kumar2018spie}.

\section{SYSTEM DYNAMICS}

\subsection{Bead dynamics}
The equation of motion of the trapped bead is described, in one dimension, by the overdamped Langevin equation
\begin{align}
    \gamma \dot{x}(t) =\underbrace{- \kappa \left(x(t)-\lambda(t)\right)}_{\text{restoring force}} +\underbrace{\sqrt{2k_{\rm{B}}T\gamma}\ \nu(t)}_{\text{thermal noise}}\ ,
\label{eq:EQM_full}
\end{align}
where $x(t)$ is the position of the bead at time \textit{t}, $\lambda(t)$ the trap center, $\gamma$ the friction coefficient, and $\nu(t)$ is Gaussian white noise, with $\langle \nu(t) \rangle = 0$ and $\langle \nu(t)\nu(t') \rangle = \delta(t-t')$. The bead measurement and feedback occur at discrete time intervals $\ts$. By integrating the dynamics over the interval $\ts$ and accounting for the relaxation of the bead during the time interval, we can derive a discrete-time equation of motion\cite{kumar2018}
\begin{align}
    x_{n+1} = \e^{-\ts/\tre}x_n + (1-\e^{-\ts/\tre})\lambda_n + \sqrt{1-\e^{-2\ts/\tre}} \zeta_n,
\end{align}
where $x_n\equiv x(n\ts)$ is the position at the $n^\mrm{th}$ time step, $\tre = \gamma/\kappa$ the trap relaxation time, and $\zeta_n$ is a Gaussian random variable with zero mean and unit variance. That is, $\langle \zeta_n \rangle = 0$, and $\langle \zeta_m\zeta_n \rangle = \delta_{mn}$.

\subsection{Trap dynamics}

\begin{figure} [ht]
   \begin{center}
   \includegraphics[width = 0.55\textwidth]{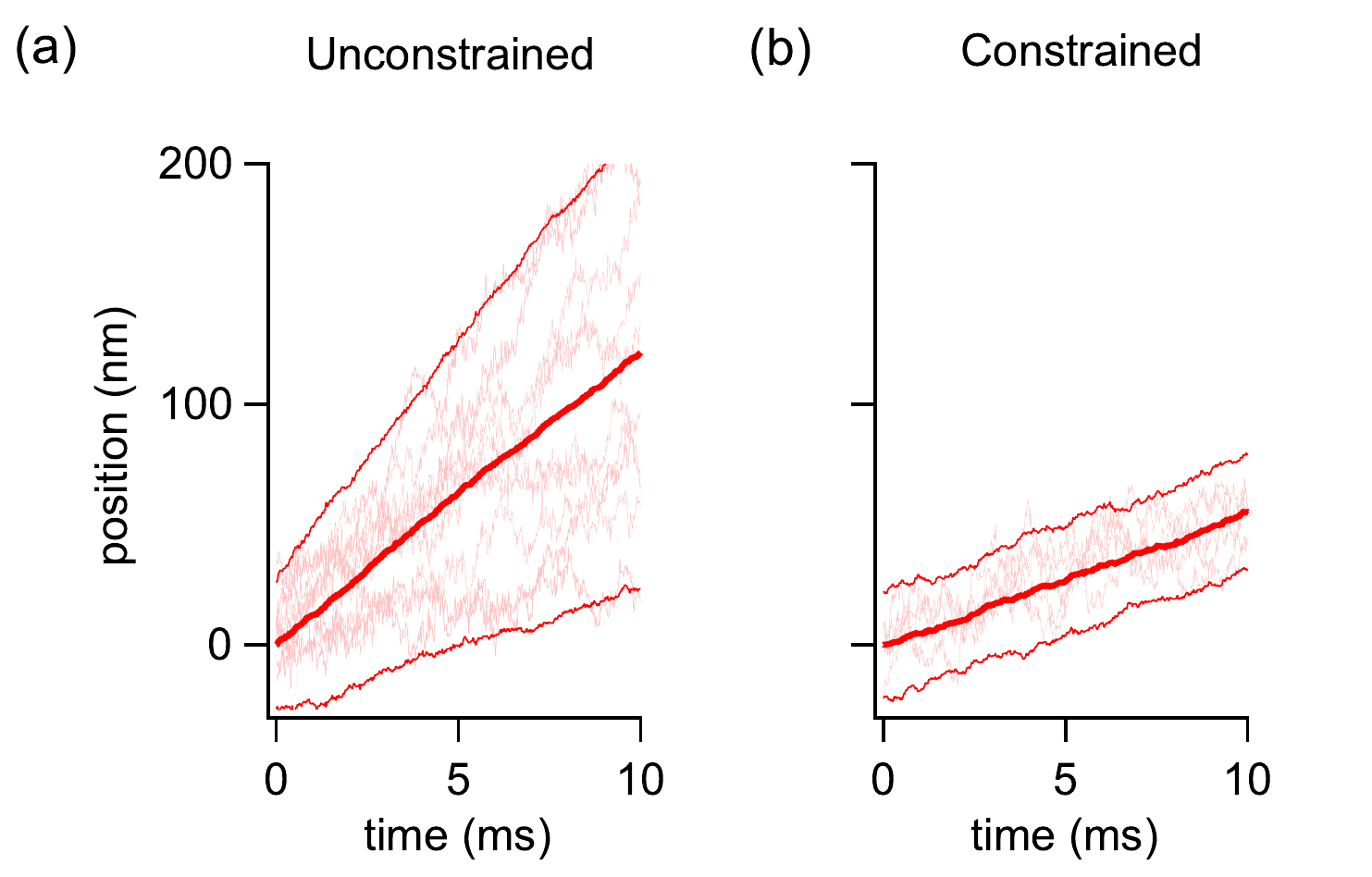}
   \end{center}
   \caption{ \label{fig:triangle} 
Bead trajectories for constrained and unconstrained feedback algorithms. (a) Mean bead position (solid red line shows average over 184 trajectories) for unconstrained feedback algorithm, with ten representative trajectories. (b) Mean bead position (solid red line shows average over 132 trajectories) for the tracking feedback algorithm, with five representative trajectories. Thin solid red lines represent two standard deviations of the bead trajectories in (a), (b). The data in (a) is from Ref.~\citeonline{saha2021}.}
\end{figure} 

\noindent As described earlier, the bead is propelled in the medium by the operation of the information engine, which exploits favorable fluctuations. Since these fluctuations occur randomly in time, beads in different trials follow different paths. Figure \ref{fig:triangle}a shows the bead positions (light red lines) when using the previously described algorithm, $\lambda_{n+1} = \lambda_n+\alpha(x_n-\lambda_n)$ when $x_n>\lambda_n$; else,  $\lambda_{n+1} = \lambda_n$. \cite{saha2021} For a fixed diameter and trap stiffness, the bead moves at an average, but fluctuating velocity (red solid line). Note that individual trajectories diverge from the average; the thin red lines represent two standard deviations of the bead trajectories. To restrict the divergence of trajectories with time, we introduce a \textit{tracking feedback control}, given by
\begin{align}
    \lambda_{n+1} =
    \begin{cases}
         \lambda_n + \alpha (x _{n} -\lambda_{n})\,, &\quad r_n>x_n>\lambda_n \,,\, \text{or }\; r_n< x_n <\lambda_n \,,\\
         \lambda_n\,, & \quad \mrm{otherwise} \,,
    \end{cases}
    \label{eq:feedback}
\end{align}
where $\alpha$ is the feedback gain. The feedback algorithm depends on the desired reference position $r_n$ and can track it in either direction. In the previous algorithm, the bead could ratchet only in one direction.

The feedback gain is tuned to $\alpha=1.7$, which ensures that the average input work is zero~\cite{saha2021}. Figure \ref{fig:triangle}b shows the average position (solid red line) of the bead and five individual trajectories (light red lines). The trajectories do not diverge with time. The desired velocity in the constrained algorithm was chosen to be slower than the average velocity obtained in the unconstrained case -- hence the smaller slope in Fig.~\ref{fig:triangle}b. The information engine cannot track velocities faster than the unconstrained feedback algorithm.

\begin{figure} [ht]
   \begin{center}
   \includegraphics[width = 0.8\textwidth]{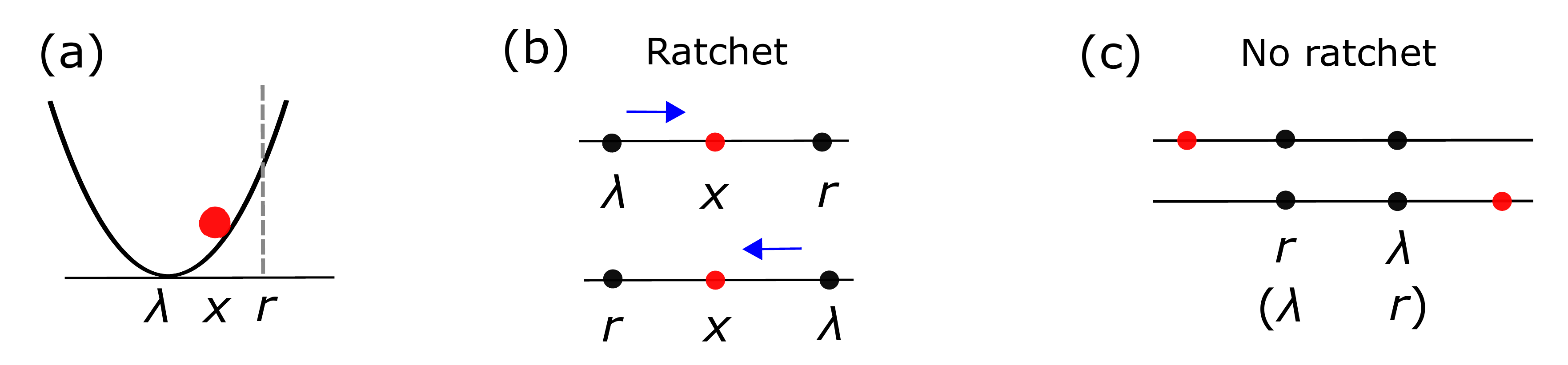}
   \end{center}
   \caption{ \label{fig:Feedback} Schematic of the feedback algorithm. (a) Trap position ($\lambda$), bead position ($x$) and reference ($r$) for a trapped bead. (b)  Trap position ratchets when $\lambda<x<r$ or when $r<x<\lambda$.  (c)  No ratcheting for other combinations of $r$ and $\lambda$, where the red circles represent the bead positions.
}
\end{figure}

A schematic diagram of the feedback algorithm is shown in Fig.~\ref{fig:Feedback}. To understand it, we note that if, after a position measurement, the trap force is already in the ``right'' direction (i.e., that it will push the particle towards the reference), then  no action is taken by the controller.  But for forces in the ``wrong'' direction (away from the reference), the controller ratchets to correct the bead position.  Notice that the algorithm ratchets when the bead position is measured to be between the reference and the trap. The algorithm thus captures all fluctuations that push the trap towards the reference signal. Some fluctuations overshoot the reference or push the bead in the opposite direction. In such cases, the trap position stays constant until the next favorable fluctuation. By neglecting the unfavorable fluctuations, the feedback rule constrains the position variance to a narrow band (thin red lines), as shown in Fig. \ref{fig:triangle}b. The trap stiffness for all the experiments is 40 pN/\textmu m, and the corresponding average velocity of the unconstrained information engine is $v_\mrm{max} = 12$ \textmu m/s\cite{saha2021}. We also tried a feedback algorithm that ratcheted after every measurement, so as to always push the particle towards the reference. However, we found the $\alpha$ values required for zero input work led to persistent oscillation of the bead about the desired trajectory.

\section{Trajectory control}
\label{sec:traj_Contrl}

\begin{figure} [ht]
   \begin{center}
   \includegraphics[width = 0.9\textwidth]{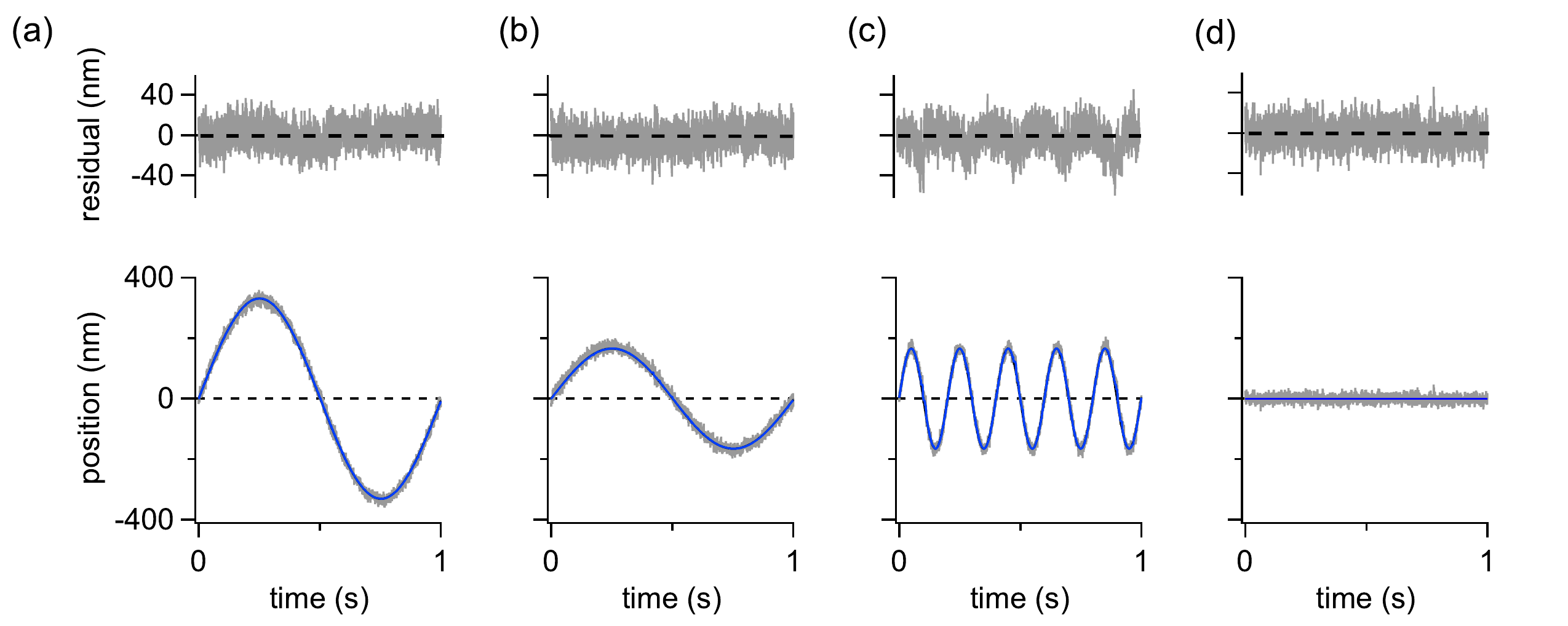}
   \end{center}
   \caption{ \label{fig:sine} 
Controlled trajectory of the bead to follow sine waves. The bead position ($x$, gray) and trap position ($\lambda$, blue) for different reference ($r$, black) signals, and the corresponding residual. The plot represents $r(t) =$ (a) $2A\, \text{sin}(\omega t)$, (b) $A\, \text{sin} (\omega t)$, (c) $A \, \text{sin}(5\omega t)$, and (d) $r(t) = \lambda(t) = 0$, where $A=165$ nm and $\omega = 2\pi\times1$ Hz.}
\end{figure}

The average velocity of an information engine depends on the trap stiffness and bead diameter\cite{saha2021}. Therefore, a bead powered by an information engine cannot track a reference signal that requires a higher velocity than that allowed by the thermal fluctuations. Figure \ref{fig:sine} shows the trajectory of the bead (gray) for the desired sine wave reference (black). We can change both the amplitude and the frequency of the sine wave that is followed by the bead. We also show the error signal (residual), $r-x$, for each of the sine waves. For comparison, the position of a bead in a static trap is shown in Fig. \ref{fig:sine}d. This is equivalent to $r(t) = 0$, with $\lambda(t=0) = 0$ as the trap does not ratchet in this case.

\begin{figure} [ht]
   \begin{center}
   \includegraphics[width = 0.75\textwidth]{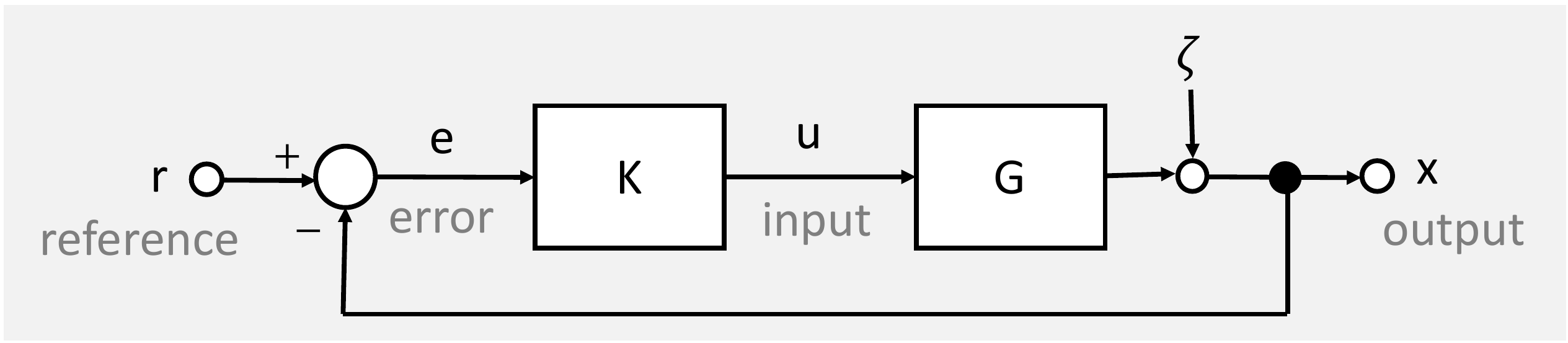}
   \end{center}
   \caption{ \label{fig:feedbackloop} 
Block diagram of the feedback loop. K denotes the controller, G the system, and $\zeta$ the thermal noise.}
\end{figure}

Figures \ref{fig:triangle}b and \ref{fig:sine} show example trajectories where the reference waves are well tracked. In all these cases, the magnitude of the velocity of the reference, $|\dot{r}(t)|$, is always below the maximum unconstrained velocity, $v_\mrm{max}$. However, the algorithm cannot respond to a reference that changes arbitrarily quickly, even when $|\dot{r}(t)| < v_\mrm{max}$. To quantify the range of frequency the feedback algorithm can track, we measure the transfer function of the controller.\cite{Bechhoefer2021} The schematic block diagram of the feedback loop is shown in Fig.~\ref{fig:feedbackloop}. Based on the error $r_n-x_n$, the controller $K$ sends an input to the system $G$. The controller is the feedback algorithm in Eq.~\ref{eq:feedback}, and the input is the force due to the trap on the bead. The corresponding output response $x$, the bead position, is measured. In addition to the reference input, there is also an input to the system that arises from the thermal noise $\zeta$. The closed-loop response of this system is then given by
\begin{subequations}
    \begin{align}
    x &= \frac{KG}{1+KG}\,r+\frac{1}{1+KG}\,\zeta\\[10pt]
    &\equiv T\,r+S\,\zeta
\end{align}
\end{subequations}
where $S$ and $T$ are the sensitivity and complementary sensitivity functions, respectively.\cite{Bechhoefer2021} Assuming that thermal fluctuations are always smaller than the reference at each reference frequency $f$, we can neglect the contribution of $S$. The closed-loop transfer function is then given by
\begin{align}
    T(f) = \frac{x(f)}{r(f)}\,,
\end{align}
where $x(f)$ and $r(f)$ are the Fourier transforms of the bead position $x(t)$ and reference $r(t)$, respectively.

\begin{figure} [ht]
   \begin{center}
   \includegraphics[width = 0.75\textwidth]{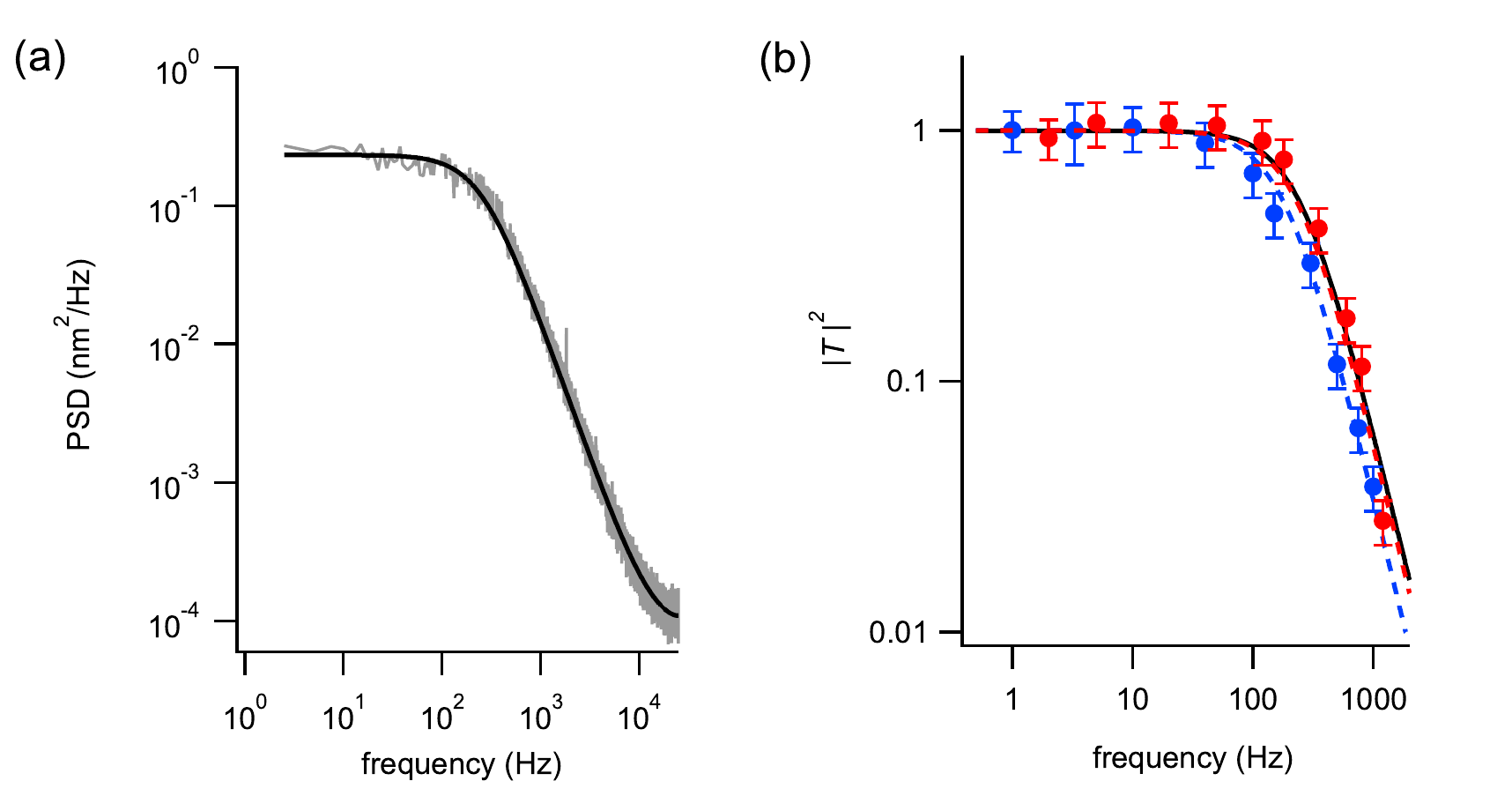}
   \end{center}
   \caption{ \label{fig:transfer} 
Transfer functions. (a) Thermal noise power spectrum density of a trapped bead (gray) calculated from a 20-s time series. Black solid line shows fit to an aliased Lorentzian\cite{berg2004} with the corner frequency $f_\mrm{c} = 254\pm2$ Hz. (b) Experimental data for $|T|^2$ for different reference frequency for information engine (blue solid markers) and open-loop (red solid markers). Blue and red dashed lines are fit to Eq.~\ref{eq:lowpass} with  $f_0 = 186\pm 11$ Hz for the information engine and $f_0^\mrm{OL} = 240\pm 13$ Hz for direct driving by sinusoidal reference inputs. Black solid line in (b) is the amplitude-normalized fit in (a).}
\end{figure}

To experimentally measure the transfer function, we input reference sine waves. For each frequency, the amplitude of the reference sine wave is chosen so that it is always below the maximum achievable velocity at the given frequency. In particular, for an input reference $r(t) = A\sin{\omega t}$, with $\omega = 2\pi f$, the amplitude $A$ is chosen to be below $v_\mrm{max}/\omega$. We record the reference signal and the corresponding bead trajectory and then calculate the power spectrum density of the bead position and reference signal. Figure \ref{fig:transfer}a shows the power spectrum density of the trapped bead, $|x(f)|^2$, when it is driven by thermal fluctuations from the surrounding water bath. 

\section{Measurement of the feedback bandwidth}

Figure~\ref{fig:transfer}b shows the reference-bead transfer function, defined as the ratio of the power of the bead and the reference as a function of reference frequency. We find that the bead follows the reference trajectory at low frequencies but not at high frequencies (blue solid markers). The feedback algorithm thus acts as a low-pass filter. After 1000 Hz, the power of the bead response to the reference goes below the thermal noise and cannot be measured. The measured frequency response is well-fit by a first-order low-pass filter~\cite{Bechhoefer2021}
\begin{align}
    |T|^2 = \left|\frac{x(f)}{r(f)}\right|^2 = \frac{1}{1+(f/f_0)^2},
    \label{eq:lowpass}
\end{align}
where $f_0$ is the frequency bandwidth of the algorithm. The DC gain (numerator) is 1, as the reference and bead position are measured in the same units. The fit to the experimental data is shown as the blue dashed line in Fig.~\ref{fig:transfer}b. For comparison, we measure the transfer function of the trapped bead (red solid markers) by driving the trap with the desired reference sine wave without feedback i.e., in open-loop. We fit the data to Eq.~\ref{eq:lowpass}, shown as red dashed line, to find the corresponding bandwidth $f_0^{\mrm{OL}}$. We find that the bandwidth $f_0 = 186\pm 11$~Hz of the feedback algorithm is slightly lower than the bandwidth $f_0^{\mrm{OL}} = 240\pm 13$~Hz of the open-loop driving. Naively, one would expect that a conventional open-loop driving, without any force constraints, would track the reference significantly better than the force-constrained information engine. However, we find that the  open-loop driving has a bandwidth comparable to that of an information engine. 

The amplitude-normalized power spectrum found from the fit is also shown in Fig.~\ref{fig:transfer}b (black solid line). Note that the open-loop bandwidth $f_0^{\mrm{OL}} = 240\pm 13$ Hz and corner frequency $f_\mrm{c} = 254 \pm 2$ Hz of the trap match well, as they are equivalent methods to measure the bandwidth. Since the bead response, when driven by an information engine, will be limited by the thermal fluctuations in the system, a close match between the controller bandwidth $f_0 = 186\pm 11$ Hz and the corner frequency $f_\mrm{c}$ suggests that any other feedback algorithm will only marginally improve the tracking performance. Thus, the information engine's response seems limited by the rate of thermal fluctuations in the system, and not by the feedback algorithm. The amplitude and frequency of the input sine wave in Fig.~\ref{fig:transfer}b are given in Table~\ref{tab:expt_amp}.

\begin{figure} [ht]
   \begin{center}
   \includegraphics[width = 0.75\textwidth]{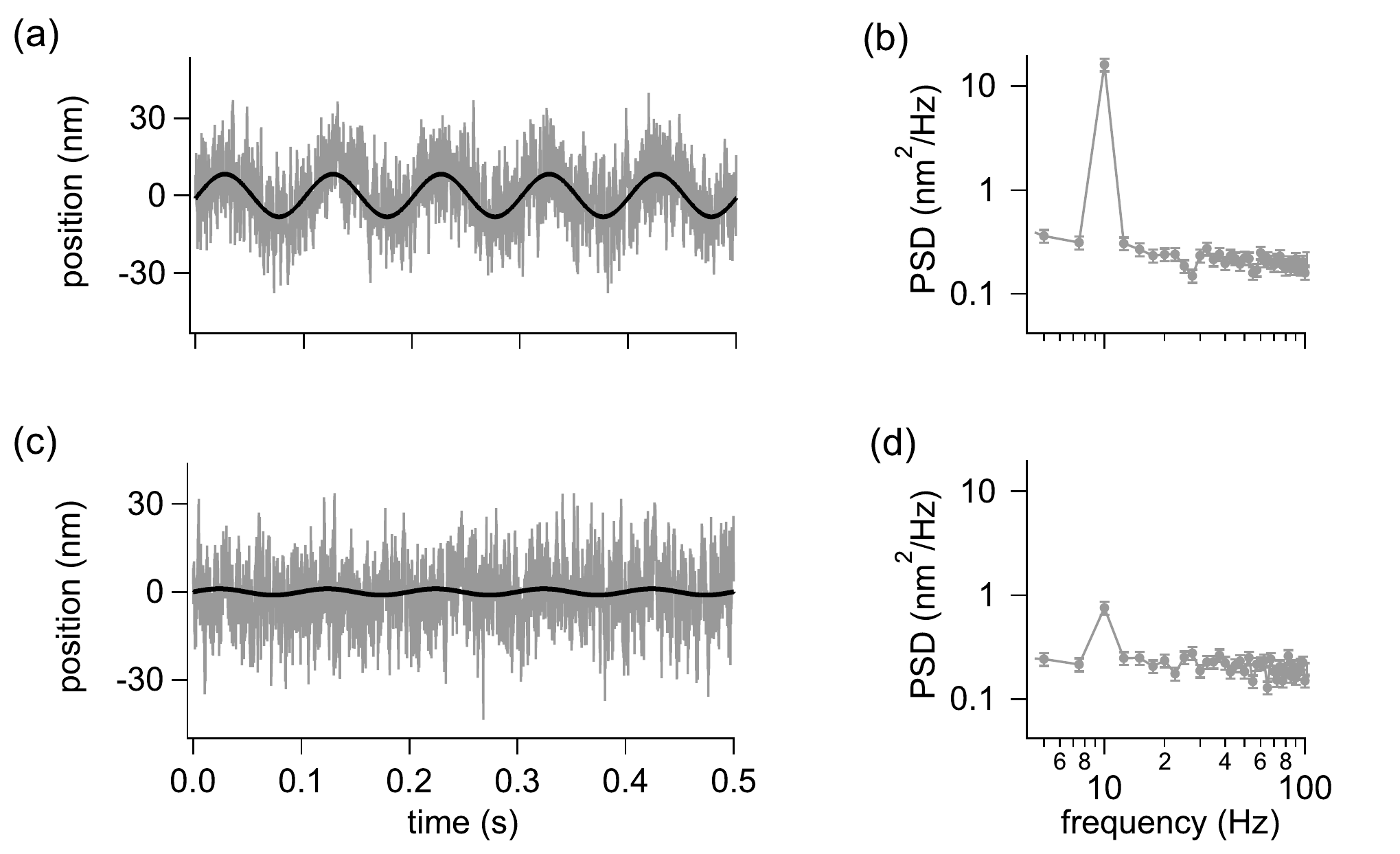}
   \end{center}
   \caption{ \label{fig:nmsine} 
Trajectory control to follow a sine wave. The bead position (gray) for different reference sine waves (black) of amplitude (a) 8 nm and (c) 1 nm at a frequency of 10 Hz. The standard deviation of the bead in the trap $\sigma = 10$ nm. (b), (d) show the corresponding power spectra calculated from time series of duration 20 s.
}
\end{figure} 

With the feedback algorithm in Eq.~\ref{eq:feedback}, the bead could follow a sine wave of amplitude 8 nm. The amplitude is comparable to the standard deviation of the bead in the trap, $\sigma = 10$ nm. The bead trajectory (gray) and the reference (black) are shown in Fig.~\ref{fig:nmsine}a. The corresponding power spectrum density shown in Fig.~\ref{fig:nmsine}b, has a peak at 10 Hz which corresponds to the frequency of the reference sine wave. We found that the bead could also track a sine wave with amplitude 1 nm, an order of magnitude smaller than $\sigma$, at 10 Hz (Fig.~\ref{fig:nmsine}c). As the amplitude is smaller than $\sigma$, one cannot see in Fig.~\ref{fig:nmsine}c that the bead does in fact follow the reference signal. However, a peak at 10 Hz in the corresponding power spectrum density (Fig.~\ref{fig:nmsine}d), confirms that the bead follows the reference signal. Of course, tracking motion at such a low amplitude is possible only because the desired trajectory is periodic, which allows averaging that reduces the effect of noise. We would expect to be able to  track a non-repeating waveform only if its amplitude exceeds the thermal noise level, $\sigma$.

\section{Conclusion}

We have developed a feedback algorithm that can constrain a bead to follow a desired trajectory, with zero input trap work. The feedback algorithm continuously ratchets when the bead lags the reference wave in either direction and stops when it reaches the reference. The algorithm is reminiscent of the bang-bang algorithm from control theory\cite{Bechhoefer2021} where the controller applies zero or maximum force in either direction to control the dynamics. It might be possible to further improve the tracking bandwidth, but we do not expect a significant increase, as the bandwidth $f_0$ of our algorithm is already close to the corner frequency $f_\mrm{c}$ of the trap. Although conventional feedback algorithms can reduce the position variance of the bead below the trap standard deviation $\sigma$, the constraint that the algorithm not allow any work to be done on the particle is a severe one, and it is not clear whether reducing the variance of fluctuations around a reference to values below that of the natural trap is possible for an information-fueled engine.

Controlled manipulation of microscopic systems has potential application in drug delivery.\cite{yoo2011bio} A major challenge for controlling the trajectory of microscopic systems is the thermal fluctuations that push them away from the target. On the contrary, information engines can convert these fluctuations into ``useful" motion. With our feedback algorithm,  an information engine can control the bead trajectory and limit the position variance to a width comparable to $\sigma$. This feature could help transport a bead to a specific location, with a precision of $\sigma$, at a time defined to a precision $\tau_\mrm{R}$. For our experiment, we recall that $\sigma=10$ nm and $\tau_\mrm{R} = 0.6$ ms. Experiments have also shown that carefully designed feedback algorithms can reduce the dissipation in stochastic systems.\cite{tafoya2019}. Integrating the ability to efficiently use thermal fluctuations and reducing thermal dissipation could also further improve the performance of current micro-bots.\cite{yoo2011bio}

\acknowledgments 
We thank Jannik Ehrich for useful comments on the manuscript.  This research was supported by grant number FQXi-IAF19-02 from the Foundational Questions Institute Fund, a donor-advised fund of the Silicon Valley Community Foundation.

\appendix
\section{Experimental parameters}
\begin{table}[!htb]
    \centering
 \begin{tabular}{|c|c|c|c|c|c|c|c|c|c|c|c|} 
    \hline
    Information engine & Frequency (Hz) & 1 & 3.3 & 10 & 40 & 100 & 150 & 300 & 500 & 750 & 1000\\
    \hline
      & Amplitude (nm) & 165 & 165 & 165 & 45.4 & 17.3 & 13.2 & 4.9 & 3.3 & 2.0 & 1.7\\
    \hline
    Open-loop & Frequency (Hz) & 2 & 5 & 20 & 50 & 120 & 180 & 350 & 600 & 800 & 1200\\
    \hline
     & Amplitude (nm) & 165 & 165 & 165 & 165 & 165 & 165 & 165 & 82.5 & 82.5 & 82.5\\
    \hline
\end{tabular}\\[10pt]
    \caption{Amplitude and frequency of the reference wave for the information ratchet and open-loop driving.}
   \label{tab:expt_amp}
\end{table}


\bibliographystyle{spiebib} 

\begin{thebibliography}{10}

\bibitem{knott1911}
Knott, C.~G.,  [{\em Life and Scientific Work of Peter Guthrie
  Tait}{\nolinebreak\hspace{0.1em}]}, Vol.~1, Cambridge Univ. Press, London
  (1911).

\bibitem{parrondo2015thermodynamics}
Parrondo, J.~M., Horowitz, J.~M., and Sagawa, T., ``Thermodynamics of
  information,'' {\em Nat. Phys.}~{\bf 11},  131--139 (2015).

\bibitem{leff2002maxwell}
Leff, H. and Rex, A.~F.,  [{\em Maxwell's {D}emon 2: {E}ntropy, {C}lassical and
  {Q}uantum {I}nformation, {C}omputing}{\nolinebreak\hspace{0.1em}]}, CRC Press
  (2002).

\bibitem{toyabe2010}
Toyabe, S., Sagawa, T., Ueda, M., Muneyuki, E., and Sano, M., ``Experimental
  demonstration of information-to-energy conversion and validation of the
  generalized {J}arzynski equality,'' {\em Nat. Phys.}~{\bf 6},  988--992
  (2010).

\bibitem{cottet2017}
Cottet, N., Jezouin, S., Bretheau, L., Campagne-Ibarcq, P., Ficheux, Q.,
  Anders, J., Auff{\`e}ves, A., Azouit, R., Rouchon, P., and Huard, B.,
  ``Observing a quantum {M}axwell demon at work,'' {\em Proc. Natl. Acad. Sci.
  U.S.A}~{\bf 114},  7561--7564 (2017).

\bibitem{camati2016}
Camati, P.~A., Peterson, J.~P., Batalhao, T.~B., Micadei, K., Souza, A.~M.,
  Sarthour, R.~S., Oliveira, I.~S., and Serra, R.~M., ``Experimental
  rectification of entropy production by {M}axwell’s demon in a quantum
  system,'' {\em Phys. Rev. Lett.}~{\bf 117},  240502 (2016).

\bibitem{koski2015chip}
Koski, J.~V., Kutvonen, A., Khaymovich, I.~M., Ala-Nissila, T., and Pekola,
  J.~P., ``On-chip {M}axwell’s demon as an information-powered
  refrigerator,'' {\em Phys. Rev. Lett.}~{\bf 115},  260602 (2015).

\bibitem{masuyama2018}
Masuyama, Y., Funo, K., Murashita, Y., Noguchi, A., Kono, S., Tabuchi, Y.,
  Yamazaki, R., Ueda, M., and Nakamura, Y., ``Information-to-work conversion by
  {M}axwell’s demon in a superconducting circuit quantum electrodynamical
  system,'' {\em Nat. Commun.}~{\bf 9},  1--6 (2018).

\bibitem{chida2017power}
Chida, K., Desai, S., Nishiguchi, K., and Fujiwara, A., ``Power generator
  driven by {M}axwell’s demon,'' {\em Nat. Commun.}~{\bf 8},  1--7 (2017).

\bibitem{kumar2018sorting}
Kumar, A., Wu, T.-Y., Giraldo, F., and Weiss, D.~S., ``Sorting ultracold atoms
  in a three-dimensional optical lattice in a realization of {M}axwell’s
  demon,'' {\em Nature}~{\bf 561},  83--87 (2018).

\bibitem{thorn2008}
Thorn, J.~J., Schoene, E.~A., Li, T., and Steck, D.~A., ``Experimental
  realization of an optical one-way barrier for neutral atoms,'' {\em Phys.
  Rev. Lett.}~{\bf 100},  240407 (2008).

\bibitem{vidrighin2016}
Vidrighin, M.~D., Dahlsten, O., Barbieri, M., Kim, M., Vedral, V., and
  Walmsley, I.~A., ``Photonic {M}axwell’s demon,'' {\em Phys. Rev.
  Lett.}~{\bf 116},  050401 (2016).

\bibitem{koski2014}
Koski, J.~V., Maisi, V.~F., Pekola, J.~P., and Averin, D.~V., ``Experimental
  realization of a {S}zilard engine with a single electron,'' {\em Proc. Natl.
  Acad. Sci. U.S.A}~{\bf 111},  13786--13789 (2014).

\bibitem{berut2012}
B{\'e}rut, A., Arakelyan, A., Petrosyan, A., Ciliberto, S., Dillenschneider,
  R., and Lutz, E., ``Experimental verification of {L}andauer’s principle
  linking information and thermodynamics,'' {\em Nature}~{\bf 483},  187--189
  (2012).

\bibitem{jun2014}
Jun, Y., Gavrilov, M., and Bechhoefer, J., ``High-precision test of
  {L}andauer’s principle in a feedback trap,'' {\em Phys. Rev. Lett.}~{\bf
  113},  190601 (2014).

\bibitem{orlov2012}
Orlov, A.~O., Lent, C.~S., Thorpe, C.~C., Boechler, G.~P., and Snider, G.~L.,
  ``Experimental test of {L}andauer's principle at the sub-$k_\text{{B}}{T}$
  level,'' {\em Jpn. J. Appl. Phys}~{\bf 51},  06FE10 (2012).

\bibitem{talukdar2017}
Talukdar, S., Bhaban, S., and Salapaka, M.~V., ``Memory erasure using
  time-multiplexed potentials,'' {\em Phys. Rev. E}~{\bf 95},  062121 (2017).

\bibitem{saha2021}
Saha, T.~K., Lucero, J.~N., Ehrich, J., Sivak, D.~A., and Bechhoefer, J.,
  ``Maximizing power and velocity of an information engine,'' {\em Proc. Natl.
  Acad. Sci. U.S.A}~{\bf 118},  e2023356118 (2021).

\bibitem{Ritort2019}
Ribezzi-Crivellari, M. and Ritort, F., ``Large work extraction and the
  {L}andauer limit in a continuous {M}axwell demon,'' {\em Nat. Phys.}~{\bf
  15},  660--664 (2019).

\bibitem{paneru2018}
Paneru, G., Lee, D.~Y., Tlusty, T., and Pak, H.~K., ``Lossless {B}rownian
  information engine,'' {\em Phys. Rev. Lett.}~{\bf 120},  020601 (2018).

\bibitem{admon2018}
Admon, T., Rahav, S., and Roichman, Y., ``Experimental realization of an
  information machine with tunable temporal correlations,'' {\em Phys. Rev.
  Lett.}~{\bf 121},  180601 (2018).

\bibitem{lopez2008}
Lopez, B.~J., Kuwada, N.~J., Craig, E.~M., Long, B.~R., and Linke, H.,
  ``Realization of a feedback controlled flashing ratchet,'' {\em Phys. Rev.
  Lett.}~{\bf 101},  220601 (2008).

\bibitem{lau2020electron}
Lau, B. and Kedem, O., ``Electron ratchets: {S}tate of the field and future
  challenges,'' {\em J. Chem. Phys.}~{\bf 152},  200901 (2020).

\bibitem{skaug2018}
Skaug, M.~J., Schwemmer, C., Fringes, S., Rawlings, C.~D., and Knoll, A.~W.,
  ``Nanofluidic rocking {B}rownian motors,'' {\em Science}~{\bf 359},
  1505--1508 (2018).

\bibitem{serreli2007}
Serreli, V., Lee, C.-F., Kay, E.~R., and Leigh, D.~A., ``A molecular
  information ratchet,'' {\em Nature}~{\bf 445},  523--527 (2007).

\bibitem{ragazzon2018}
Ragazzon, G. and Prins, L.~J., ``Energy consumption in chemical fuel-driven
  self-assembly,'' {\em Nat. Nanotechnol.}~{\bf 13},  882--889 (2018).

\bibitem{alvarez2008}
Alvarez-P{\'e}rez, M., Goldup, S.~M., Leigh, D.~A., and Slawin, A.~M., ``A
  chemically-driven molecular information ratchet,'' {\em J. Am. Chem.
  Soc.}~{\bf 130},  1836--1838 (2008).

\bibitem{chicos2014}
Bregulla, A.~P., Yang, H., and Cichos, F., ``Stochastic localization of
  microswimmers by photon nudging,'' {\em ACS Nano}~{\bf 8},  6542--6550
  (2014).

\bibitem{kumar2019fluid}
Kumar, D., Shenoy, A., Li, S., and Schroeder, C.~M., ``Orientation control and
  nonlinear trajectory tracking of colloidal particles using microfluidics,''
  {\em Phys. Rev. Fluids}~{\bf 4},  114203 (2019).

\bibitem{carlsen2014}
Carlsen, R.~W., Edwards, M.~R., Zhuang, J., Pacoret, C., and Sitti, M.,
  ``Magnetic steering control of multi-cellular bio-hybrid microswimmers,''
  {\em Lab Chip}~{\bf 14},  3850--3859 (2014).

\bibitem{Bechhoefer2021}
Bechhoefer, J.,  [{\em Control Theory for
  Physicists}{\nolinebreak\hspace{0.1em}]}, Cambridge Univ. Press (2021).

\bibitem{kumar2018spie}
Kumar, A. and Bechhoefer, J., ``Optical feedback tweezers,'' in [{\em Optical
  Trapping and Optical Micromanipulation XV}{\nolinebreak\hspace{0.1em}]},
  {\bf 10723},  107232J, SPIE (2018).

\bibitem{kumar2018}
Kumar, A. and Bechhoefer, J., ``Nanoscale virtual potentials using optical
  tweezers,'' {\em Appl. Phys. Lett.}~{\bf 113},  183702 (2018).

\bibitem{berg2004}
Berg-S{\o}rensen, K. and Flyvbjerg, H., ``Power spectrum analysis for optical
  tweezers,'' {\em Rev. Sci. Instum.}~{\bf 75},  594--612 (2004).

\bibitem{yoo2011bio}
Yoo, J.-W., Irvine, D.~J., Discher, D.~E., and Mitragotri, S., ``Bio-inspired,
  bioengineered and biomimetic drug delivery carriers,'' {\em Nat. Rev. Drug
  Discov.}~{\bf 10},  521--535 (2011).

\bibitem{tafoya2019}
Tafoya, S., Large, S.~J., Liu, S., Bustamante, C., and Sivak, D.~A., ``Using a
  system’s equilibrium behavior to reduce its energy dissipation in
  nonequilibrium processes,'' {\em Proc. Natl. Acad. Sci. U.S.A}~{\bf 116},
  5920--5924 (2019).

\end{thebibliography}

\end{document}